\input harvmac

%\draftmode
{\Title{\vbox{
\hbox{CALT-68-2373, CITUSC/02-006}
\hbox{\tt hep-th/0202157}}}
{\vbox{
\centerline{Penrose Limit of ${\cal N}=1$ Gauge Theories}}}
\vskip .3in
\centerline{Jaume Gomis and Hirosi Ooguri}
\vskip .4in
}

\centerline{ California Institute of Technology 452-48,
Pasadena, CA 91125}

\vskip .4in

We find a Penrose limit of $AdS_5\times T^{1,1}$ which 
gives the pp-wave geometry identical to the one that
appears in the Penrose limit of $AdS_5\times S^5$. 
This leads us to conjecture that there is 
a subsector of the corresponding ${\cal N}=1$ gauge theory 
which has enhanced ${\cal N}=4$ supersymmetry. 
We identify operators in the ${\cal N}=1$ gauge theory 
with stringy excitations in the pp-wave geometry and discuss how 
the gauge theory
operators fall into ${\cal N}=4$ supersymmetry multiplets.
We find similar enhancement of symmetry in some other models, 
but there are also examples in which there is no supersymmetry 
enhancement in the Penrose limit.

\vfill
\eject
\newsec{Introduction}

The AdS/CFT correspondence 
%\MaldacenaRE
\lref\MaldacenaRE{
J.~Maldacena,
``The large $N$ limit of superconformal field theories and supergravity,''
Adv.\ Theor.\ Math.\ Phys.\  {\bf 2} (1998) 231,
{\tt arXiv:hep-th/9711200}.}
%%CITATION = HEP-TH 9711200;%%
%\GubserBC
\lref\GubserBC{
S.~S.~Gubser, I.~R.~Klebanov and A.~M.~Polyakov,
``Gauge theory correlators from non-critical string theory,''
Phys.\ Lett.\ B {\bf 428}(1998) 105,
{\tt arXiv:hep-th/9802109}.}
%%CITATION = HEP-TH 9802109;%%
%\WittenQJ
\lref\WittenQJ{
E.~Witten,
``Anti-de Sitter space and holography,''
Adv.\ Theor.\ Math.\ Phys.\  {\bf 2} (1998) 253,
{\tt arXiv:hep-th/9802150}.}
%%CITATION = HEP-TH 9802150;%%
\refs{\MaldacenaRE ,\GubserBC ,\WittenQJ} relates
a conformal field theory in $(p+1)$-dimensions to
string theory in $AdS_{p+2}\times X$, where $X$ is a compact Einstein space.
 In 
the last few years, we have learned much  
about nonperturbative aspects of
string theory and conformal field theories
using this correspondence 
%\AharonyTI
\lref\AharonyTI{
O.~Aharony, S.~S.~Gubser, J.~Maldacena, H.~Ooguri and Y.~Oz,
``Large N field theories, string theory and gravity,''
Phys.\ Rept.\  {\bf 323} (2000) 183,
{\tt arXiv:hep-th/9905111}.}
%%CITATION = HEP-TH 9905111;%%
\AharonyTI . One of the major 
obstructions in making further
progress in this direction has been our lack
of understanding of the worldsheet dynamics describing
string theory in $AdS$ and related backgrounds. 
Understanding this problem is essential, for example,
to finding quantitative results from string theory for the large $N$
limit of  
gauge
theories with finite 't Hooft coupling.
Although much progress has
been made for string theory in $AdS_3$ with an NS-NS background
%\MaldacenaHW
\lref\MaldacenaHW{
J.~Maldacena and H.~Ooguri,
``Strings in $AdS_3$ and $SL(2,R)$ WZW model. Part 1: The Spectrum,''
J.\ Math.\ Phys.\  {\bf 42} (2001) 2929,
{\tt arXiv:hep-th/0001053};
%%CITATION = HEP-TH 0001053;%%
%\MaldacenaKV
J.~Maldacena, H.~Ooguri and J.~Son,
``Strings in $AdS_3$ and the $SL(2,R)$ WZW model. Part 2: Euclidean black
hole,''
J.\ Math.\ Phys.\  {\bf 42} (2001) 2961
{\tt arXiv:hep-th/0005183};
%%CITATION = HEP-TH 0005183;%%
%\MaldacenaKM
J.~Maldacena and H.~Ooguri,
``Strings in $AdS_3$ and the $SL(2,R)$ WZW model. Part 3: Correlation
functions,''
{\tt arXiv:hep-th/0111180}; and references therein.}
%%CITATION = HEP-TH 0111180;%%
\MaldacenaHW , 
worldsheet dynamics in higher 
dimensional $AdS$ spaces and/or with R-R field strengths
remains a mystery. Thus far, most of the results obtained from string
theory in  $AdS$ 
have relied on the supergravity approximation. 

Recently, it was shown in 
%\MetsaevBJ
\lref\MetsaevBJ{
R.~R.~Metsaev,
``Type IIB Green-Schwarz superstring in plane wave Ramond-Ramond
background,'' 
{\tt arXiv:hep-th/0112044}.}
%%CITATION = HEP-TH 0112044;%%
\MetsaevBJ\ that the worldsheet theory of the Type IIB string on the
maximally supersymmetric pp-wave geometry 
%\BlauNE
\lref\BlauNE{
M.~Blau, J.~Figueroa-O'Farrill, C.~Hull and G.~Papadopoulos,
``A new maximally supersymmetric background of IIB superstring theory,''
JHEP {\bf 0201}, 047 (2002),
{\tt arXiv:hep-th/0110242}.}
%%CITATION = HEP-TH 0110242;%%
%\BlauDY
\lref\BlauDY{
M.~Blau, J.~Figueroa-O'Farrill, C.~Hull and G.~Papadopoulos,
``Penrose limits and maximal supersymmetry,''
{\tt arXiv:hep-th/0201081}.}
%%CITATION = HEP-TH 0201081;%%
\BlauNE :
\eqn\ppmetric{
 ds^2 = -4dx^+dx^- + \sum_{i=1}^8 \left( dr^idr^i - r^ir^i dx^+dx^+
\right)
,}
with constant R-R 5-form flux,
\eqn\constf{ F_{+1234}=F_{+5678} =\hbox{const},}
is exactly soluble 
in the light-cone Green-Schwarz formalism
%\GreenWT
\lref\GreenWT{
M.~B.~Green and J.~H.~Schwarz,
``Covariant description of superstrings,''
Phys.\ Lett.\ B {\bf 136} (1984) 367.}
%%CITATION = PHLTA,B136,367%%
%\GreenSG
\lref\GreenSG{
M.~B.~Green and J.~H.~Schwarz,
``Properties of the covariant formulation of superstring theories,''
Nucl.\ Phys.\ B {\bf 243} (1984) 285.}
%%CITATION = NUPHA,B243,285;%%
\refs{\GreenWT,\GreenSG}. Moreover, in a recent interesting paper 
%\BerensteinJQ
\lref\BerensteinJQ{
D.~Berenstein, J.~Maldacena and H.~Nastase,
``Strings in flat space and pp waves from ${\cal N}$ = 4 super Yang Mills,''
{\tt arXiv:hep-th/0202021}.}
%%CITATION = HEP-TH 0202021;%%
\BerensteinJQ , it was pointed out that
Type IIB string theory on this
pp-wave background is
dual to the large $N$ limit of a certain 
subsector of four dimensional ${\cal N}=4$ 
$SU(N)$ supersymmetric gauge theory. 
The subsector is characterized
by choosing a $U(1)_R$ subgroup of the $SU(4)_R$
R-symmetry of the gauge theory
and by considering states with 
conformal weight $\Delta$
and $U(1)_R$ charge $R$ which scale as $\Delta, R
\sim \sqrt{N}$  and whose difference
$(\Delta - R)$ is finite in the large $N$ limit. The claim 
is that, in the $N\rightarrow \infty$ 
limit with the string coupling $g_s = g_{YM}^2$ finite, the subspace
of the gauge theory Hilbert space and the operator algebra preserving
these conditions are described by  string
theory in  the pp-wave geometry. This duality was
derived in \BerensteinJQ\ by starting with the familiar correspondence
between ${\cal N}=4$ $SU(N)$ supersymmetric gauge theory
 and Type IIB string theory in
$AdS_5\times S^5$ and considering a scaling 
limit of  the  geometry
near a null geodesic   
in  $AdS_5\times S^5$ carrying large angular momentum with respect
to the $U(1)_R$ isometry of $S^5$. This corresponds
to truncating to the appropriate subsector of the gauge theory
in the scaling limit. The string theory 
background that one obtains in the scaling limit is the pp-wave
geometry with a constant R-R flux
which can then be quantized in the light-cone
gauge \MetsaevBJ . The scaling limit is a special example
of the Penrose limit which transforms any solution
of supergravity 
 to a plane wave geometry \lref\penrose{R. Penrose, ``Any
space-time has a plane wave as a limit,''
in {\sl Differential geometry and relativity,}
pp. 271-275, Reidel, Dordrecht, 1976.}
\lref\guven{R. G\"uven, ``Plane wave limits and
T-duality,'' {\sl Phys. Lett.} B482 (2000) 255,
{\tt arXiv: hep-th/0005061}.}\lref\bfp{
M. Blau, J.Figueroa-O'Farrill, and G.Papadopoulos,
``Penrose limits, supergravity and brane dynamics,''
{\tt arXiv: hep-th/0202111}.}
\refs{\penrose,\guven,\BlauDY,\bfp}.

 In \BerensteinJQ , it was shown
that operators in the appropriate subsector of
${\cal N}=4$ $SU(N)$ gauge theory
 can be identified with ${\it stringy}$
oscillators in the pp-wave background. 
This matching makes quantitative predictions about the 
spectrum of the gauge theory beyond the supergravity
approximation, and some of them were checked in \BerensteinJQ\
using gauge theory computation of  planar Feynman diagrams. 

In this paper we consider a similar duality that exists between a
certain four-dimensional ${\cal N}=1$ gauge theory
and Type IIB string theory in a pp-wave background. 
%\KlebanovHH
\lref\KlebanovHH{
I.~R.~Klebanov and E.~Witten,
``Superconformal field theory on three-branes at a Calabi-Yau
singularity,'' 
Nucl.\ Phys.\ B {\bf 536} (1998) 199,
{\tt arXiv:hep-th/9807080}.}
%%CITATION = HEP-TH 9807080;%
We derive this duality by taking a scaling limit of
the duality \KlebanovHH\
between Type IIB string theory on $AdS_5\times T^{1,1}$
and the four-dimensional superconformal field theory which
consists of an ${\cal N}=1$ $SU(N)\times SU(N)$ super Yang-Mills
multiplet with a pair of bifundamental chiral multiplets $A_i$ and $B_i$
transforming in the $(N, \bar{N})$  and  $(\bar{N}, N)$ representation
of the gauge group. The gauge theory is
flown to the IR fixed point and deformed by
an $SU(2)_1\times SU(2)_2$ invariant
superpotential
\eqn\potential{ W = {\lambda \over 2} \epsilon^{ij}\epsilon^{i'j'}
\Tr A_i B_{i'} A_j B_{j'} ,}
which is exactly marginal at the fixed point.
This gives the 
theory of \KlebanovHH\ that lives\foot{ The details of the gauge theory
will appear in section  
$3$.}  on  $N$ D3-branes sitting at the conifold
singularity of a Calabi-Yau three-fold. The scaling limit is obtained
by considering the geometry near a null geodesic carrying large
angular momentum in the $U(1)_R$ isometry of the $T^{1,1}$ space which
is dual to the $U(1)_R$ R-symmetry in the ${\cal N}=1$
superconformal field theory. 

The scaling limit around this null geodesic in $AdS_5\times T^{1,1}$
results in a pp-wave 
geometry. We identify the light-cone
Hamiltonian, longitudinal momentum and angular momentum of string
theory in the pp-wave  
geometry with linear combinations of the conformal weight
 $\Delta$, $U(1)_R$ charge $R$  and  $U(1)_1\times U(1)_2$ global
charges $Q_1$ and $Q_2$ 
of
 operators in the  dual ${\cal N}=1$ gauge
theory. 
The appropriate scaling limit requires truncating the gauge
theory Hilbert space to those operators whose conformal weight
$\Delta$, R-charge $R$  and global charges $Q_1$ and $Q_2$ 
scaling like 
\eqn\scalinglike{\Delta,~ R, ~Q_1, ~Q_2\sim \sqrt{N},}
with
\eqn\scalingwith{
\Delta-{3\over 2}R ,~ Q_1-{1\over 2}R,~ {\rm and}~~ Q_2-{1\over 2}R
:~{\rm finite},} 
in the large $N$ limit.  

Remarkably, the pp-wave geometry that one obtains in the scaling limit
can be transformed into the maximally supersymmetric background
of \ppmetric\ and \constf\ after a suitable change of coordinates. Therefore,
in this limit supersymmetry is enhanced. This implies that the
subsector of the Hilbert space  
of the ${\cal N}=1$ gauge theory of \KlebanovHH\ obeying
the conditions \scalinglike\ and \scalingwith\ has
a hidden ${\cal N}=4$ supersymmetry. We believe that
this is a very interesting prediction of our duality that deserves
further study. 

We find that the change of coordinates induces 
{\it twisting} of the light-cone
Hamiltonian of the string theory by
\eqn\stringtwist{ p^-=p^-_{S^5}+ J_1 + J_2,}
where $p^-_{S^5}$ is the Hamiltonian of the maximally supersymmetric wave
found in \MetsaevBJ\ and $J_1$, $J_2$ correspond
to rotation charges under an ${\bf R}^2 \times {\bf R}^2$ subspace
of the transverse space of the pp-wave geometry.  
From the gauge theory point of view, the light-cone
Hamiltonian $p^-$ before the twisting is $\Delta - {3 \over 2} R$
and the rotational charges are given by $J_a = Q_a - {1\over 2} R$ 
($a=1,2$). Note that they remain finite in the limit
\scalingwith .
After the twisting, the light-cone Hamiltonian
is identified with $\Delta_{{\cal N}=4} - R_{{\cal N}=4}$,
in terms of the conformal weight and the R charge
for the ${\cal N}=4$ supersymmetry algebra. 
Thus we find the following relation
\eqn\gaugetwist{ 
\eqalign{\Delta_{{\cal N}=4} - R_{{\cal N}=4} & =
\Delta - {3\over 2} R - J_1 - J_2\cr
&= \Delta - {1\over 2} R - Q_1 - Q_2.}}
The spectrum of stringy excitations in the ${\cal N}=4$ theory 
studied in \BerensteinJQ\ can then be turned into that of 
the ${\cal N}=1$ theory by this twisting. The twisted
string spectrum is highly degenerate, and we show that it
matches with gauge theory expectations. 

The Penrose limit focuses on geometry near a null geodesic. 
When we have a gauge theory whose supersymmetry is reduced
by placing branes on a curved space, the Penrose limit may
flatten out the space and restore  supersymmetry. We have found
other examples where similar enhancement of symmetry takes place.
Those include the ${\cal N}=1$ pure super Yang-Mills theory
(with Kaluza-Klein tower of fields) 
studied in 
%\MaldacenaYY
\lref\MaldacenaYY{
J.~M.~Maldacena and C.~Nu\~nez,
``Towards the large $N$ limit of pure ${\cal N}$
 = 1 super Yang Mills,''
Phys.\ Rev.\ Lett.\  {\bf 86}, 588 (2001),
{\tt arXiv:hep-th/0008001}.}
%%CITATION = HEP-TH 0008001;%%
\MaldacenaYY\ and gauge theories realized on 
branes on a $C^3/Z_3$ orbifold singularity. 
The limit of the former is a variation of 
the Nappi-Witten geometry
%\NappiIE
\lref\NappiIE{
C.~R.~Nappi and E.~Witten,
``A WZW model based on a nonsemisimple group,''
Phys.\ Rev.\ Lett.\  {\bf 71}, 3751 (1993),
{\tt arXiv:hep-th/9310112}.
%%CITATION = HEP-TH 9310112;%%
}
\NappiIE\ with 16 supercharges, and that of the
latter is the maximally supersymmetric pp-wave.
On the other hand, there are
cases in which such enhancement does not happen, such
as gauge theories on branes at a $C^2/Z_2$ orbifold singularity.

The rest of the paper is organized as follows. In section 2, 
we consider the scaling limit around a null geodesic in $AdS_5 \times
T^{1,1}$ and show that one obtains a pp-wave background. We identify
the subsector of the Hilbert space of the  dual superconformal field
theory that is dual to string theory in the pp-wave
geometry. We show that there is a coordinate transformation which
brings the pp-wave background to the one which has maximal supersymmetry
\ppmetric . The Hamiltonian of the pp-wave is then obtained by
twisting  the Hamiltonian of the Type IIB string in
that background by angular momentum charges $J_1$ and $J_2$ that the
strings carry in the maximally supersymmetric pp-wave background.
In section $3$, we describe ingredients
of the ${\cal N}=1$ theory of \KlebanovHH\ that are required to
compare the gauge theory and the string theory. Precise
matching is obtained by identifying in a specific way
string theory excitations with gauge theory operators. 
In section 4, we discuss other examples in which similar enhancement
of symmetry takes place and show an example where symmetry
enhancement does not occur.
 We conclude
with a discussion. In the appendix we explicitly solve for the
worldsheet theory of the pp-wave background that we obtain in the
limit. Explicit diagonalization of the Hamiltonian shows that it is
related to that of the maximally supersymmetric wave by twisting.

\bigskip
\noindent
{\bf Note added:}

After posting this paper on the e-Print arXiv, we have
received %\ItzhakiKH
\lref\ItzhakiKH{
N.~Itzhaki, I.~R.~Klebanov and S.~Mukhi,
``PP wave limit and enhanced supersymmetry in gauge theories,''
arXiv:hep-th/0202153.
%%CITATION = HEP-TH 0202153;%%
}
%\ZayasRX
\lref\ZayasRX{
L.~A.~Zayas and J.~Sonnenschein,
``On Penrose limits and gauge theories,''
arXiv:hep-th/0202186.
%%CITATION = HEP-TH 0202186;%%
}
\refs{\ItzhakiKH, \ZayasRX}, where the Penrose limit
of $AdS_5 \times T^{1,1}$ is studied and the
supersymmetry enhancement is also noted. We also received
%\RussoRQ
\lref\RussoRQ{
J.~G.~Russo and A.~A.~Tseytlin,
``On solvable models of type IIB superstring in NS-NS and R-R plane wave  backgrounds,''
arXiv:hep-th/0202179.
%%CITATION = HEP-TH 0202179;%%
} \RussoRQ , 
where the Penrose limit of backgrounds with NS-NS 3 form field
and its relation to a generalization of the Nappi-Witten
model are also discussed.

\newsec{Penrose limit of $AdS_5\times T^{1,1}$}

We start by considering the supergravity solution dual to the  ${\cal
N}=1$ superconformal field theory of \KlebanovHH\  that we
describe in the next section. The background of interest is
$AdS_5\times T^{1,1}$, where $T^{1,1}=(SU(2)\times SU(2))/U(1)$,
with the $U(1)$ diagonally embedded in the two $SU(2)$'s. The
Einstein metric on $AdS_5\times 
T^{1,1}$ is given by
\eqn\metrica{\eqalign{
ds^2_{Ads}&=L^2\left(-dt^2\hbox{cosh}^2\rho+d\rho^2+\hbox{sinh}^2\rho
d\Omega_3\right)\cr
ds^2_{T^{1,1}}&=L^2\left({1\over
9}(d\psi+\hbox{cos}\ \theta_1d\phi_1+\hbox{cos}\ \theta_2d\phi_2)^2
\right. \cr&~~~~~~ \left. +{1\over
6}(d\theta_1^2+\hbox{sin}^2\theta_1d\phi_1^2+d\theta_2^2+
\hbox{sin}^2\theta_2d\phi_2^2)\right),}}
where $d\Omega_3$ is the volume form of a unit $S^3$
and the curvature radius $L$ of $AdS_5$ is given by $L^4=4\pi
g_{s}N{\alpha^\prime}^2 27/16$. Topologically, $T^{1,1}$ is a $U(1)$
bundle over ${\bf S}^2\times {\bf S}^2$. The base is parametrized by
coordinates 
$(\theta_1,\phi_1)$  and $(\theta_2,\phi_2)$ respectively and the Hopf
fiber coordinate $\psi$ has period $4\pi$. The $SU(2)_1\times
SU(2)_2\times U(1)_R$ isometry group of $T^{1,1}$ is identified 
with the $SU(2)_1\times SU(2)_2$ global symmetry and $U(1)_R$ symmetry
of the dual superconformal field theory of \KlebanovHH . 
In addition, the solution has a constant dilaton and a R-R five-form flux
\eqn\flux{
F=L^4(\hbox{vol}_{AdS}+\hbox{vol}_{T^{1,1}}),}
where $\hbox{vol}_{AdS}, \hbox{vol}_{T^{1,1}}$ are the volume forms of
$AdS_5$ and $T^{1,1}$.

We now perform a scaling limit around a null geodesic in $AdS_5\times
T^{1,1}$ which rotates along the $\psi$ coordinate of $T^{1,1}$, whose
shift symmetry corresponds to the $U(1)_R$ symmetry of the dual
superconformal field theory\foot{Shifts along the angles $\phi_1$ and
$\phi_2$ generate an $U(1)\times U(1)$ subgroup of the $SU(2)_1\times
SU(2)_2$ isometries and correspond in the gauge theory side
 to the abelian charges $Q_1$ 
and $Q_2$, which are the Cartan generators of the $SU(2)_1\times
SU(2)_2$ global symmetry group of the gauge theory.}. We
introduce coordinates which label the geodesic
\eqn\lightcone{\eqalign{
x^+&={1\over 2}\left(t+{1\over 3}(\psi+\phi_1+\phi_2)\right)\cr
x^-&={L^2\over 2}\left(t-{1\over 3}(\psi+\phi_1+\phi_2)\right).}}
and consider a scaling limit around $\rho=\theta_1=\theta_2=0$ in the
geometry \metrica . We
take $L \rightarrow \infty$  while rescaling the coordinates
\eqn\rescale{
\rho={r\over L}\qquad \theta_1={\sqrt{6}\over L}\xi_1 \qquad
\theta_2={\sqrt{6}\over L}\xi_2.} 
The metric one obtains in the limit is
\eqn\ppwave{\eqalign{
ds^2=&-4dx^+dx^- +\sum_{i=1}^4\left( dr^i dr^i  -r^i r^i dx^+dx^+\right) \cr
& +\sum_{a=1,2} \left( d\xi_a^2+\xi_a^2d\phi_a^2
- 2\xi_a^2d\phi_a dx^+ \right)  \cr
=&-4dx^+dx^- +\sum_{i=1}^4\left( dr^i dr^i  -r^i r^i dx^+dx^+\right) \cr
& + \sum_{a=1,2} \left[ dz_ad\bar{z}_a + i(\bar{z}_a dz_a - z_a d{\bar
z}_a)dx^+\right] .}} 
In the last line, we introduced  complex Cartesian 
coordinates $z_a$ 
in lieu of $(\xi_a, \phi_a)$.  The metric has a covariantly constant
null Killing vector $\partial/\partial x^-$ so that it is a pp-wave metric.
The pp-wave has a natural decomposition of the ${\bf R}^8$ transverse space 
into ${\bf R}^4\times {\bf 
R}^2\times {\bf  R}^2$, where ${\bf R}^4$ is parametrized by $r^i$ and ${\bf 
R}^2\times {\bf  R}^2$ by $z_a$. The  geometry is supported by a null,
covariantly 
constant flux of the R-R field,
\eqn\fluxpp{
F_{+1234}=F_{+5678}= \hbox{const}.}
The obvious symmetries of this background are the $SO(4)$ rotations in
${\bf R}^4$ and a $U(1)\times U(1)$ symmetry\foot{In fact,
the metric and the flux is invariant under a
larger symmetry as we shall see below.} 
rotating ${\bf R}^2\times {\bf  R}^2$. 
In the gauge theory side, the $SO(4)$ symmetry 
corresponds to the subgroup of the $SO(2,4)$ conformal symmetry
($i.e.$ the rotations of $S^3$  in the field theory space 
${\bf R} \times S^3$) and the $U(1) \times U(1)$ rotation charges
$J_1$ and $J_2$ with the  $U(1)\times U(1)$ symmetry $Q_1-{1\over 2}R$
and $Q_2-{1\over 2}R$ respectively, where $R$ is the $U(1)_R$ charge of the
gauge theory and $Q_1$, $Q_2$ are 
the Cartan generators 
 of the $SU(2)_1 \times SU(2)_2$ global symmetry of the
dual superconformal field theory. 

In order to compare string theory in the pp-wave
geometry \ppwave\  with the appropriate subsector of the dual field
theory determined by the limit, \lightcone\ and \rescale ,
we need to establish the correspondence
between conserved charges in string theory
and in  gauge theory. In string theory, the light-cone
momenta can be identified with combinations of the conformal
weight $\Delta$ and the $U(1)_R$ charge $R$ of the dual
superconformal field theory by noting that\foot{The factor of two in
the normalization of $R$ 
is due to the $4\pi$ periodicity of the $\psi$ coordinate.}
\eqn\charges{\eqalign{
2p^-&=i\partial_{x^+}=i(\partial_t+3 \partial_\psi)=\Delta - {3\over
2}R\cr
2p^+&={i\over L^2} \partial_{x^-}={i\over L^2} (\partial_t-3 
\partial_\psi)= {1\over L^2}\left(\Delta + {3\over 2} R\right).}}
The $J_1$ and $J_2$ rotation charges of the string can be identified with 
\eqn\chargesglob{
J_a=-i{\partial\over \partial \phi_a}_{\big|x^\pm}=-i{\partial\over 
\partial \phi_a}_{\big|t,\psi}
+i{\partial \over \partial\psi}_{\big|t,\phi_i}=
Q_a-{1\over 2}R\qquad 
a=1,2}
such that the states of the dual gauge theory are also labeled by
these 
global symmetry charges.

Therefore, it follows from the identification \charges\ and the
$L\rightarrow \infty$ limit, \lightcone\ and \rescale , 
that string theory in
the pp-wave background \ppwave\ with finite $p^-,p^+$ and $J_i$ 
is dual to the ${\cal N}=1$ gauge theory of
\KlebanovHH\ in a subsector of the Hilbert space where $\Delta,R, Q_a
 \sim L^2\sim
\sqrt{N}$ with finite $(\Delta-{3\over 2}R)$ and $Q_a-{1\over 2}R$ 
in the large $N$
limit. In 
particular the duality with string theory predicts that there is a  set
of non-chiral primary operators, which satisfy $\Delta >{3\over 2}R$, 
whose dimension and R-charge grow without bound but such that the deviation
from the BPS bound is finite in the large $N$ limit. In the next section
we will make a proposal for which operators of the gauge theory obey
this peculiar scaling behavior.

Remarkably, the pp-wave geometry \ppwave\ that we have obtained in
the scaling limit reduces to the maximally supersymmetric pp-wave
solution \ppmetric\ 
 after performing a coordinate dependent $U(1)\times U(1)$ rotation
in the ${\bf R}^2\times {\bf R}^2$ plane as
\eqn\rot{\eqalign{
z_a&=e^{ix^+}w_a\cr
\bar{z}_a&=e^{-ix^+}\bar{w}_a.}}
This means that the symmetry and supersymmetry of the original
$AdS_5\times T^{1,1}$ background is maximally enhanced in the 
Penrose limit,
\lightcone\ and \rescale . We interpret this as saying
 that the corresponding
subsector of the dual ${\cal N}=1$ superconformal field theory has
 hidden ${\cal N}=4$ supersymmetry.

The coordinate transformation \rot\ mapping the solution
\ppwave\ to the maximally supersymmetric solution \ppmetric\
allows us to write down the string Hamiltonian $p^-$ in \charges\ in
terms of the Hamiltonian $p^-_{S^5}$ of the maximally
symmetric solution already computed in \MetsaevBJ . 
Using the coordinate transformation \rot\ and the relation \charges\
between the isometry and the gauge theory charges, we find that 
\eqn\hamiltwist{\eqalign{\Delta - {3\over 2}R & =
2p^- \cr
& =i{\partial \over \partial x^+}_{\big| z_a} \cr
& =i{\partial \over \partial x^+}_{\big| w_a} 
+\sum_a\left(w_a{\partial \over \partial w_a}-{\bar
w}_a{\partial \over \partial {\bar w}_a}\right)\cr
&=2p^-_{S^5}+J_1+J_2, }}
where $J_1$ and $J_2$ are the U(1) rotation charges around an ${\bf
R}^2\times {\bf R}^2$ subspace of the maximally supersymmetric pp-wave
${\bf
R}^8$ transverse space\foot{The $U(1)\times U(1)$ rotation charges
when changing from 
$z$ to $w$ coordinates remain identical.}, and
\eqn\hamilbefore{2p^-_{S^5} =\Delta_{{\cal N}=4}-R_{{\cal N}=4}.}

We now briefly recall the string spectrum $p^-_{S^5}$ found in
\MetsaevBJ . The spectrum consists of a set of eight bosonic harmonic
oscillators $a^i_n$ and eight fermionic  harmonic
oscillators $S^\alpha_n$ with $i,\alpha=1,2,\ldots 8$. The
increase in light-cone energy due to one oscillator is given by
\eqn\contri{
2\delta p^-_{S^5}=\sqrt{1+\left({n\over \alpha^\prime p^+}\right)^2},}
so that in particular the zero modes  $a^i_0$ and $S^a_0$ increase
$p^-_{S^5}$ by one. To obtain the spectrum of $p^-$ we use the
twisting formula \hamiltwist. In order to find the
spectrum we need to know the charges of $a^i_n$ and $S^\alpha_n$ under
the $U(1)\times U(1)$ subgroup of the $SO(8)$ rotation group.

The charges of the bosonic oscillators follow from decomposing the
bosonic oscillators, which transform under the ${\bf 8}_v$
representation of the $SO(8)$ rotation group under $SU(2)\times
SU(2)\times U(1)\times U(1)$, such that the ${\bf R}^8$ space on
which $SO(8)$ acts splits as ${\bf R}^4\times{\bf R}^2\times{\bf
R}^2$, and $ U(1)\times U(1)$ rotates ${\bf R}^2\times{\bf
R}^2$. We organize the oscillators as 
\eqn\oscill{\eqalign{
a^i_n&\qquad J_1=J_2=0\qquad i=1,2,3,4\cr
w^1_n&\qquad J_1=1, J_2=0\cr
\bar{w}^1_n&\qquad J_1=-1, J_2=0\cr
w^2_n&\qquad J_1=0, J_2=1\cr
\bar{w}^2_n&\qquad J_1=0, J_2=-1.}}
Therefore, the contribution to $p^-$ of each of the bosonic
oscillators is
\eqn\contri{\eqalign{
a^i_n&\qquad 2\delta p^-=\sqrt{1+\left({n\over \alpha^\prime p^+}\right)^2}\cr
w^1_n&\qquad  2\delta p^-=\sqrt{1+\left({n\over \alpha^\prime
p^+}\right)^2}+1\cr 
\bar{w}^1_n&\qquad  2\delta p^-=\sqrt{1+\left({n\over \alpha^\prime
p^+}\right)^2}-1\cr 
w^2_n&\qquad 2\delta p^-=\sqrt{1+\left({n\over \alpha^\prime
p^+}\right)^2}+1\cr
\bar{w}^2_n&\qquad 2\delta p^-=\sqrt{1+\left({n\over \alpha^\prime
p^+}\right)^2}-1.}}
The fermionic oscillator contribution to $p^-$ follows by looking at the
$U(1)\times U(1)$ charges carried by the $SO(8)$ spinor ${\bf 8}_s$
under  $SU(2)\times
SU(2)\times U(1)\times U(1)$. The oscillators split as
\eqn\chargesdecom{
{\bf 8}_s\rightarrow ({\bf 2},{\bf 1})_{(1/2,1/2)}\oplus  ({\bf 2},{\bf
1})_{(-1/2,-1/2)}\oplus    
({\bf 1},{\bf 2})_{(1/2,-1/2)}\oplus  
({\bf 1},{\bf 2})_{(-1/2,1/2)},} 
where the charges in the subscript correspond to $(J_1,J_2)$ charges.
Therefore, their contribution to the
light-cone Hamiltonian \hamiltwist\ is 
\eqn\contriferm{\eqalign{
S^{\alpha ++}_n&\qquad 2\delta p^-=\sqrt{1+\left({n\over \alpha^\prime
p^+}\right)^2}+1\cr 
S^{\alpha --}_n&\qquad  2\delta p^-=\sqrt{1+\left({n\over \alpha^\prime
p^+}\right)^2}-1\cr 
S^{{\dot\alpha} +-}_n&\qquad  2\delta p^-=\sqrt{1+\left({n\over \alpha^\prime
p^+}\right)^2}\cr 
S^{{\dot\alpha} -+}_n&\qquad 2\delta p^-=\sqrt{1+\left({n\over \alpha^\prime
p^+}\right)^2}.}}
Thus we see that the spectrum of bosonic oscillators and fermionic
oscillators are identical.

We note that the light-cone energy is not increased by the action of
the bosonic zero mode oscillators $\bar{w}^1_0$ and
$\bar{w}^2_0$ nor by  the action of their  supersymmetric fermionic
zero mode partners 
$S^{\alpha --}_0$. Therefore, the 
system has an infinitely degenerate spectrum labeled by the number of
times the vacuum state is acted on by the zero modes. 
In the
original coordinates of the pp-wave in \ppwave , the degeneracy of the
spectrum can be easily understood by considering the zero-mode sector
($i.e.$, the point particle limit) of string theory.
In the zero mode sector, the Hamiltonian contains, on top of four free
harmonic oscillators, two decoupled Landau Hamiltonians describing an
electron in a magnetic field in the planes $z_1$ and $z_2$. The
degeneracy in these coordinates corresponds to the well known Landau
level degeneracy of states of the electron where the degeneracy is
labeled by the angular momentum of the electron. 
In the appendix, we extend this to stringy excitations,
quantizing the bosonic string Hamiltonian
in these coordinates. We show that indeed the spectrum is twisted 
by $(J_1+J_2)$ as in \contri\ with
respect to the maximally supersymmetric case of \MetsaevBJ. 

In the next section we
give a precise prescription of how to realize the infinite degeneracy
of states in string theory, which corresponds in the dual
superconformal field theory to having an infinite degeneracy of operators
with a given conformal dimension, and identify the operators dual to the
insertion of the string theory oscillators.

\newsec{Gauge theory spectrum}

Type IIB superstring theory in $AdS_5 \times T^{1,1}$ is dual
to the ${\cal N}=1$ gauge theory with gauge group
$SU(N)\times SU(N)$ with two chiral multiplets $A_i$ ($i=1,2$)
transforming
in the $(N,\bar{N})$ representation of the gauge group and two chiral
multiplets $B_{i'}$ ($i'=1,2$) in the 
$({\bar N},N)$ representation. The theory is flown to the
IR fixed point. We then turn on the 
superpotential \potential\
involving the chiral superfields $A_i$ and $B_{i'}$. 
At the fixed point, these chiral superfields have
conformal weight $3/4$ and  R-charge $1/2$. They transform
as $({\bf 2,1})$ and $({\bf 1,2})$ under the
$SU(2)_1\times SU(2)_2$ global symmetry.

Now we are ready to compare states in the string
theory with operators in the gauge theory. We will mainly focus only on  the
bosonic excitations of the theory, denoted by $w^a_n$ and $\bar w^a_n$
in \contri .   Let us begin with the zero
mode sector of string theory, which is generated by
$\bar w_0^1$ and $\bar w_0^2$.  Since we are
dealing with the 
zero mode of the string, they are supergravity modes in $AdS_5
\times T^{1,1}$. The correspondence between supergravity modes
and gauge theory operators has been discussed extensively
in 
%\GubserVD
\lref\GubserVD{
S.~S.~Gubser,
``Einstein manifolds and conformal field theories,''
Phys.\ Rev.\ D {\bf 59} (1999) 025006,
{\tt arXiv:hep-th/9807164}.
%%CITATION = HEP-TH 9807164;%%
}
%\CeresoleZS
\lref\CeresoleZS{
A.~Ceresole, G.~Dall'Agata, R.~D'Auria and S.~Ferrara,
``Spectrum of type IIB supergravity on $AdS_5 \times T^{11}$: 
Predictions on ${\cal N}$  = 1 SCFT's,''
Phys.\ Rev.\ D {\bf 61} (2000) 066001,
{\tt arXiv:hep-th/9905226}.
%%CITATION = HEP-TH 9905226;%%
}
\refs{\GubserVD,\CeresoleZS}. It is useful to rephrase it in
the stringy terminology of the last section. We will find some
special feature of the supergravity spectrum in the Penrose
limit. Since the light-cone
Hamiltonian $2p^-$ is equal to $(\Delta -{3\over 2}R)$, the
lowest energy states of the string theory are chiral primary states.
The basic ones are of the form $\Tr (AB)^R$. Among them, 
we can identity the ground state $|0\rangle$ of the $\bar w^a_0$ 
oscillators 
with the gauge theory operator,
\eqn\ground{ |0\rangle \leftrightarrow
 \Tr\left[ (A_1 B_1)^R\right].}
Here and in the following, we
ignore normalization factors in gauge theory
operators which may depend on $N$ and $R$.
We have chosen $A_i$ and $B_{i'}$ so
that $A_1$ and $A_2$ carries $Q_1$ charge $+{1\over 2}$ and $-{1\over 2}$
and $B_1$ and $B_2$ carries $Q_2$ charge $+{1\over 2}$ and
$-{1\over 2}$ respectively.
Their $(J_1, J_2)$ charges defined by $J_a=Q_a-{1\over 2}R$
are therefore $({1\over 4}, -{1\over 4})$ and $(-{3\over 4},-{1\over 4})$
for $A_1$ and $A_2$, and $(-{1\over 4}, {1\over 4})$ and
$(-{1\over 4}, -{3\over 4})$ for $B_1$ and $B_2$. Thus the operator
${\rm Tr}\left[ (A_1B_1)^R \right]$ in \ground\ carries
$J_1 = J_2=0$, and $\Delta - {3\over 2} R =
\Delta_{{\cal N}=4}- R_{{\cal N}=4}=0$. Namely it saturates
the BPS bounds for both ${\cal N}=1$ and ${\cal N}=4$ supersymmetry
algebras. 

The operators
$\bar w^a_0$ then act on it as
\eqn\baction{ \bar w_0^1: A_1 \rightarrow A_2,
~~\bar w_0^2: B_1 \rightarrow B_2.}
In order for these to map the
chiral primary state \ground\ into another chiral primary state, their
action has to be symmetrized along all $A_1$'s and $B_1$'s
in the trace \KlebanovHH . Since the $\bar w^a_0$ oscillators
are absent in the worldsheet Hamiltonian, their action does not increase
$2p^-=\Delta-{3\over 2}R$.  This
is consistent with the fact that the
action of the $\bar w^a_0$'s in the gauge theory gives rise to
chiral primary states saturating the BPS bound of the
${\cal N}=1$ supersymmetry. On the other hand, $\Delta_{{\cal N}=4}
- R_{{\cal N}=4} = \Delta-{3\over 2} R -J_1-J_2$ is
increased by $1$ every time we act with $\bar{w}_0$ since 
$J_1+J_2=0$ for $A_1$ and $B_1$ while it is $-1$ for 
$A_2$ and $B_2$. Note that $(J_1+J_2)$ is not positive for
any of the operators. Therefore these operators satisfy
the BPS bounds for
both the ${\cal N}=1$ and the ${\cal N}=4$ supersymmetry algebras.
This gives an important consistency check of our conjecture
about the supersymmetry enhancement.

From the way they act on $A_i$ and $B_{i'}$, 
it is clear that $\bar{w}_0^1, \bar{w}_0^2$ 
and their conjugates are identified as the raising 
and lowering operators of the $SU(2)_1 \times SU(2)_2$ 
global symmetry of the gauge theory. On the string
worldsheet, they act as harmonic oscillators.
On the other hand, when acting on the gauge theory
operators ${\rm Tr}\left[ (A_1B_1)^R\right]$, they
obey the constraints $(\bar{w}_0^1)^{R+1}=0, 
(\bar{w}_0^2)^{R+1}=0$. This does not contradict with
the correspondence between string theory and
gauge theory. Since $J_a=Q_a-
{1\over 2}R$ ($a=1,2$) has to remain finite in the 
limit $R\rightarrow \infty$, only a finite number 
of $\bar w_0$'s can act on this operator and these
constraints become irrelevant. 

The oscillators $w^a_0$'s in \contri\ are more interesting. 
They change $(\Delta - {3\over 2}R)$ by $2$, thus
their action does not generate chiral primary states.
Nevertheless the resulting states should be in
the supergravity sector. Candidates for such states
can be found in the list of operators given in \CeresoleZS
, where they are called semi-conserved superfields.
Although they are not chiral primaries, their conformal
dimensions are protected. 
The ones we are interested in here take the following form,
\eqn\whatazero{
 \Tr\left[ \left(A e^V \bar{A} e^{-V}\right)^{n_1} 
\left(e^{V} \bar{B} e^{-V} B\right)^{n_2} (AB)^R \right],}
where $V$ is the vector multiplet for the gauge
group $SU(N)\times SU(N)$. There is one important subtlety
in making the identification. It was pointed out 
in \GubserVD\ that, in order for the corresponding
supergravity mode to have a rational conformal dimension,
the integers $n_1$ and $n_2$ must satisfy the
Diophantine equation,
\eqn\diophantine{n_1^2 + n_2^2 -4n_1n_2 -n_1-n_2=0.}
This is true if we are studying states with finite 
$\Delta$ and $R$. Since we are studying the
scaling limit $\Delta, R \sim \sqrt{N} \rightarrow \infty$, 
it is worth revisiting its origin. The constraint comes from the
fact that the eigenvalue $E$ of the Laplacian on $T^{1,1}$
for the corresponding supergravity mode takes the form
\eqn\laplace{ E= 6n_1^2 + 6n_2^2 + 8n_1n_2
+ (6R+8)(n_1+n_2) +{3\over 2}R\left({3\over2}R+4\right).}
One can then show that the conformal weight
$\Delta = -2 +\sqrt{4 + E}$ of the mode becomes rational
if \diophantine\ is satisfied. However, this condition
is relaxed in the limit $R \rightarrow \infty$.
In this limit, the meaningful quantity is $(\Delta - {3\over 2}R)$,
and it is given by
\eqn\whathappens{
\Delta - {3\over 2} R = 2n_1 + 2n_2+ O\left({1\over R}\right).}
The right-hand side of this equation is clearly rational. 
In fact, they are even integers.
Therefore the Diophantine constraint \diophantine\ is
irrelevant in the subsector of the Hilbert space we are looking. 
Moreover the formula is exactly what we need to identify
the action of $w^a_0$. Since each of these operators
is supposed to increase $2p^-=\Delta - {3\over 2}R$ by $2$
and they carry the $U(1)\times U(1)$ quantum numbers 
$(-1,0)$ and $(0,-1)$, the natural identification is
\eqn\whata{ w_0^1: {\rm insertion~of~}
A_1 e^V \bar{A}_2 e^{-V},~~
w_0^2 : {\rm insertion~of~}
e^{V} \bar B_2 e^{-V} B_1 .}
Of course the insertion must be symmetrized along the
trace so that $(\Delta - {3\over 2} R)$ is minimized. 

As we discussed in the last section, string theory
in $AdS_5 \times T^{1,1}$ acquires enhanced ${\cal N}=4$ superconformal
symmetry in the Penrose limit. This means that the
spectrum of the gauge theory operators in this subsector
must fall into ${\cal N}=4$ multiplets. Since the oscillators
$w^a_0$ and $\bar w^a_0$ are part of the ${\cal N}=4$ superconformal
generators ($P^I$ and $J^{+I}$ in the notation of 
%\MetsaevRE
\lref\MetsaevRE{
R.~R.~Metsaev and A.~A.~Tseytlin,
``Exactly solvable 
model of superstring in Ramond-Ramond plane wave background,''
{\tt arXiv:hep-th/0202109}.}
%%CITATION = HEP-TH 0202109;%%
\MetsaevBJ , see also \MetsaevRE),
we conjecture that the chiral primary fields
of the form $\Tr (AB)^R$ and the semi-conserved multiplets
of the form \whatazero\ combine to make ${\cal N}=4$ multiplets
in the limit.\foot{The decomposition of ${\cal N}=4$ chiral multiplets
into ${\cal N}=1$ multiplets has been discussed in 
\lref\FerraraBP{
S.~Ferrara, M.~A.~Lledo and A.~Zaffaroni,
``Born-Infeld corrections to D3 brane action in 
$AdS_5 \times S^5$ and ${\cal N}$ = 4,  $d$ = 4 primary superfields,''
Phys.\ Rev.\ D {\bf 58}, 105029 (1998)
[arXiv:hep-th/9805082].
%%CITATION = HEP-TH 9805082;%%
} \FerraraBP , whose results should be useful in proving 
this conjecture. There, the R-symmetry acting on ${\cal N}=1$
multiplets is chosen to be the commutant of $SU(3)$ in the $SU(4)$ 
R-symmetry of the ${\cal N}=4$ theory. On the other hand, 
the R-symmetry of the ${\cal N}=1$ theory of \KlebanovHH\ 
is the commutant of $SU(2) \times SU(2)$ 
in the $SU(4)$ R-symmetry of the enhanced 
${\cal N}=4$ supersymmetry. }
Note that this can happen only in the limit. For finite
$R$, the semi-conserved multiplets have to obey the
Diophantine constraint \diophantine\ in order for them
to have rational conformal weights.  

Now let us turn to the stringy excitations. 
The string-bit interpretation suggests that
a worldsheet operator with non-zero Fourier
mode $n$ acts on a gauge theory operator
just as its $n=0$ counter-part, but the
action is summed over the trace with a
position dependent phase proportional to
$n$.  This point
of view was adopted in \BerensteinJQ\ to identify operators in
in the ${\cal N}=4$ gauge theory corresponding to
stringy excitations. We extend their proposal to
the ${\cal N}=1$ theory we are studying. 

Consider the oscillators $\bar w_n^a$. When $n=0$,
it is defined as the replacement $A_1 \rightarrow A_2$
for $a=1$ and $B_1\rightarrow B_2$ for $a=2$, averaged
over the trace. It is then natural to identify
\eqn\whatbn{
\eqalign{  \bar w_n^1 |0\rangle &\leftrightarrow
  \sum_{k=0}^{R-1} \Tr \left[ (A_1B_1)^{k} A_2 B_1 (A_1B_1)^{R-1-k}
\right] e^{{2\pi i nk\over R}} \cr
 \bar w_n^2 |0\rangle &\leftrightarrow
  \sum_{k=0}^{R-1} \Tr \left[ (A_1B_1)^{k} A_1 B_2 (A_1B_1)^{R-1-k}
\right] e^{{2\pi i nk\over R}} .}}
Of course, the operators on the right-hand side vanish due
to the cyclicity of the trace. This corresponds to the
string theory fact that the left-hand side does not satisfy
the level matching momentum constraint, as explained in \BerensteinJQ . 
The idea is to use an analogous definition for
$\bar w_{n_1}^1 \cdots  \bar w_{n_s}^1
\bar w_{m_1}^2 \cdots \bar w_{m_t}^2 |0\rangle$
such that $\sum n_i + \sum m_{i'} =0$. For example,
\eqn\example{
\bar w_{-n}^1 \bar w_n^2 |0\rangle
\leftrightarrow 
\sum_{k=0}^{R-1} \Tr \left[ A_2 (B_1A_1)^k B_2 (A_1B_1)^{R-1-k}\right]
e^{{2\pi i nk\over R}}.}

Operators of this type are not chiral primaries. 
As pointed out in  \KlebanovHH , these operators vanish if we use
the constraint due to the superpotential \potential ,
\eqn\potentialconstraint{
 A_1B_{i'} A_2 =A_2 B_{i'}A_1,~~~B_1A_i B_2= B_2A_i B_1.}
Since they are defined against the potential wall, they gain
additional conformal weights beyond their naive values,
and therefore $\Delta - {3\over 2}R >0$.
The string theory computation in section 3 predicts that
the action of $\bar w_n$ changes $(\Delta -{3\over 2}R)$
by
\eqn\changeofd{ 
\eqalign{ \delta\left(\Delta -{3\over 2}R\right)_n
=& \sqrt{ 1+ \left({n \over \alpha'p^+}\right)^2}-1 \cr
=& \sqrt{1 + 3\pi g_sn^2 {N  \over R^2}}
-1.}}

We are taking the large $N$ limit so that $N/R^2$
remains finite. When $g_s > 0$, the right-hand side is indeed 
strictly positive. One of the interesting features of 
this formula is that $\delta(\Delta-{3\over 2}R)_n$
 vanishes at $g_s=0$ even for $n\neq 0$, giving rise to
further degeneracy of the spectrum. 
In the large $N$ limit, the only parameter of the string
theory is $g_s$ while that of the gauge theory
is $\lambda$, the coefficient of the superpotential.
Since the superpotential must increase the conformal
weights of these operators, one explanation 
of what happens at $g_s=0$ is that $\lambda$ vanishes
in the gauge theory side.
To our knowledge, a map between these parameters has
not been worked out, and this observation may give us 
some hint about the correspondence between the gauge
theory moduli and the string theory moduli. At $g_s = 0$, the string
theory Hilbert space becomes the Fock space of
first quantized strings, whose spectrum is integral. 
It would be interesting to
see whether such a structure emerges in this subsector
of the gauge theory at $\lambda=0$.

The other set of operators $w^a_n$ are
similarly interpreted in the gauge theory side.
These operators insert $Ae^{V}\bar A e^{-V}$
and $e^V \bar B e^{-V} B$ in the trace and
sum over insertion points along the trace with
position dependent phases. The string theory
computation predicts that the amount of change
of $(\Delta - {3\over 2}R)$ is given by
\eqn\change{ 
\eqalign{ \delta\left(\Delta -{3\over 2}R\right)_n
=& \sqrt{ 1+ \left({n \over \alpha'p^+}\right)^2}+1 \cr
=& \sqrt{1 + 3\pi g_sn^2 {N  \over R^2}}
+1.}}

One can also consider operators corresponding to
stringy excitations in the $r^i$ directions.
In \BerensteinJQ , these are interpreted as taking
derivative of operators with respect to
spatial coordinates in the gauge theory. In our
case, one may be puzzled by the fact that there
seem to be two types of derivatives, those
acting on $A$'s and those acting on $B$'s. 
Clearly only particular combinations of them
correspond to stringy excitations of the $r^i$
directions. In general, there are many gauge invariant 
observables one can write down, and only some of them 
correspond to string states. We expect that the others become
infinitely heavy, $i.e.$ $(\Delta - {3\over 2}R)$
becomes infinitely large, in the large $N$ limit.

\newsec{Other examples}

The Penrose limit focuses on the geometry near a null geodesic. 
When we have a gauge theory whose supersymmetry is reduced
by placing branes on a curved space, the Penrose limit may
flatten out the space near the branes and restore supersymmetry. 
Thus we expect that enhancement of symmetry takes place in
a large class of theories which have gravity duals. 

Let us consider the supergravity solution found in \MaldacenaYY\   
which is dual to 
 pure ${\cal N}=1$ supersymmetric 
Yang-Mills theory (with Kaluza-Klein tower of fields).
We will show that it 
has a Penrose limit which is identical to  that of a
collection of
five-branes in flat space. Thus, in this case,
symmetry is again enhanced in the corresponding 
subsector of  ${\cal N}=1$ super Yang-Mills.
To exhibit this we write the metric of NS-NS five-branes in flat space 
\eqn\metrfive{
ds^2=ds^2(R^{1,5})+L^2\left(d\rho^2+{1\over
4}(d\psi+\cos\theta\ d\phi)^2+{1\over 4}d\theta^2+
{1\over 4}\sin^2\theta\ d\phi^2\right),} 
where $L^2=\alpha^\prime N$. We introduce the following null
coordinates
\eqn\nullcoor{\eqalign{
x^+&={1\over 2}\left({t\over L}+{1\over 2}(\psi+\phi)\right)\cr
x^-&={L^2\over 2}\left({t\over L}-{1\over 2}(\psi+\phi)\right)}}
and consider the limit around $\rho=\theta=0$. We take the limit
$L\rightarrow \infty$ while rescaling the coordinates
\eqn\rescaleccor{
\rho={r\over L}\qquad \theta={2y\over L}.}
The metric that one obtains in this limit is 
\eqn\limitpen{
ds^2=-4dx^+dx^--2y^2d\phi dx^++ds^2({\bf R}^8),}
which by using the coordinate transformation in \rot\ reduces to the
pp-wave metric  
\eqn\limitpenz{
ds^2=-4dx^+dx^--w\bar{w}dx^+ dx^++ dwd\bar w+ ds^2({\bf R}^6),}
where $w$ is a complex coordinate on an ${\bf R}^2$ plane in ${\bf
R}^8$. Moreover,  in the limit \limitpen\ 
the dilaton becomes constant and the NS-NS three-form flux
$H_{+12}=\hbox{const}$  
becomes null. The worldsheet theory
describing the coordinates $(x^+,x^-,w,\bar w)$ can be identified with 
the WZW model based on the non-semi-simple group which is
the central extension of the two-dimensional Poincare group,
found by Nappi and Witten \NappiIE .

One can also show that the Penrose limit of the Maldacena-Nu\~nez solution
\MaldacenaYY\ in the region near $\rho=0$ gives rise to a
generalization of the Nappi-Witten geometry with 16 supercharges
and again the supersymmetry is enhanced.\foot{In the earlier version
of this paper, we claimed that the Penrose limit of the
Maldacena-Nu\~nez solution is identical to the Nappi-Witten
geometry \limitpenz . We thank Juan Maldacena and Horatiu
Nastase for pointing out an error in our argument. Our observation
about the enhancement of supersymmetry in this case still 
remains true.} 
It would be interesting to explore the
consequences of this symmetry enhancement for  the gauge theory.

Another interesting example is to consider the Penrose limit of the
dual pair obtained by placing $N$ D3-branes at a  $C^3/Z_3$ orbifold
singularity. The gauge theory living on the D3-branes is an ${\cal
N}=1$ four-dimensional quiver gauge theory
%\DouglasSW
\lref\DouglasSW{
M.~R.~Douglas and G.~W.~Moore,
``D-branes, quivers, and ALE instantons,''
{\tt arXiv:hep-th/9603167}.}
%%CITATION = HEP-TH 9603167;%%
\DouglasSW\ and the gravity dual is $AdS_5\times S^5/Z_3$
%\KachruYS
\lref\KachruYS{
S.~Kachru and E.~Silverstein,
``4d conformal theories and strings on orbifolds,''
Phys.\ Rev.\ Lett.\  {\bf 80}, 4855 (1998),
{\tt arXiv:hep-th/9802183}.}
%%CITATION = HEP-TH 9802183;%%
\KachruYS ,  where the $S^5$ is described by the complex coordinates
$z_i$ ($i=1,2,3$) constrained by
\eqn\sphere{
|z_1|^2+|z_2|^2+|z_3|^2=1}
and the $Z_3$ generator $g$ acts by
\eqn\orbact{
g\cdot z_i= \alpha z_i\qquad i=1,2,3\qquad \alpha^3=1,}
so that $Z_3$ acts freely on the sphere. The $U(1)_R$ symmetry of the 
gauge theory can be identified with shifts along the Hopf fiber
coordinate $\psi$ when $S^5/Z_3$ is described as a $U(1)$ bundle over
$CP^2$
\eqn\metric{
ds^2=(d\psi+\sin^2 \mu\ \sigma_3)^2+d\mu^2+\sin^2\mu\left
(\sigma_1^2+\sigma_2^2+\cos^2\mu\ \sigma_3^2\right),} 
where $\sigma_i$ are a set of left-invariant $SU(2)$ one forms
satisfying $d\sigma_i=\epsilon_{ijk}\sigma_j\wedge \sigma_k$. The
$Z_3$ orbifold group acts by restricting the range of the $\psi$
coordinate to $1/3$ of the usual value for $S^5$ while leaving all
other coordinates intact. We introduce null coordinates 
\eqn\nullorbc{\eqalign{
x^+&={1\over 2}\left(t+\psi\right)\cr
x^-&={L^2\over 2}\left({t}-\psi\right)}}
taking the $L\rightarrow \infty$ near $\mu=\rho=0$ with\foot{The
$\rho$ coordinate comes the $AdS_5$ part of the metric.}
\eqn\rescaleorb{
\rho={r\over L}\qquad \mu={y\over L}.} 
In the limit one gets 
\eqn\metrusig{
ds^2=-4dx^+dx^-+\sum_{i=1}^4
(dr^idr^i-r^ir^idx^+dx^++dy^idy^i)+2y^2dx^+\sigma_3,}
where $y_i$ are Cartesian coordinates in ${\bf R}^4$. By introducing a
pair of complex coordinates $z_a$ for  ${\bf R}^4$ one can show
that 
\metrusig\ can be rewritten as  
\eqn\metrusig{
ds^2=-4dx^+dx^-+\sum_{i=1}^4
(dr^idr^i-r^ir^idx^+dx^+)+\sum_a\left(dz_ad\bar{z}_a+
i(z_ad\bar{z}_a-\bar{z}_adz_a)dx^+\right),}
which we have already shown is the same as the maximally supersymmetric
pp-wave metric \ppmetric .
So we have another example in which a subsector of a four dimensional
${\cal N}=1$ gauge theory is enhanced to ${\cal N}=4$. It would be
interesting to match the string oscillators in this background with
operators in the quiver gauge theory.

We should note, however, that the Penrose limit does not
always enhance supersymmetry. For example, we can consider the
dual pair generated by placing a collection of D3-branes at a
$C^2/Z_2$ orbifold singularity. The gauge theory is a four dimensional
${\cal N}=2$ gauge theory 
\DouglasSW\ and the gravity dual is $AdS_5\times S^5/Z_2$
\KachruYS , where the $S^5$ is described by \sphere\ 
and the $Z_2$ generator $g$ acts by
\eqn\orbact{
g\cdot z_1= -z_1\qquad g\cdot z_2=-\ z_2.}
Thus the $Z_2$ action has as fixed locus an $S^1$ described by
$|z_3|^2=1$ at $z_1=z_2=0$. The coordinate parametrizing the fixed
$S^1$ can be identified with the $\psi$ coordinate considered
in the
Penrose limit of $AdS_5\times S^5$ \BerensteinJQ . 
Therefore, the corresponding
pp-wave limit is given by the $Z_2$ orbifold of the maximally
supersymmetric pp-wave \ppmetric, where $Z_2$ acts on an ${\bf R}^4$
subspace of the transverse  ${\bf R}^8$ space. The main difference
between this and the previous orbifold is that $\psi$ is not acted by
$Z_2$ while the other angles on the $S^5$ have $Z_2$ identifications.
This is related to the fact that the $S^5$ has a fixed circle in
this case. 

To find the amount of supesymmetry left unbroken by the orbifold, one
must find which components of the Killing spinors of the pp-wave geometry
\ppmetric\ are left invariant under the $Z_2$ action. The Killing
spinors of \ppmetric\ were found \BlauNE\ and take the form
\eqn\spinors{
\epsilon(x,y,x^+)=f(x,y,x^+)\epsilon_0,}
where $x$ and $y$ are the transverse ${\bf R}^4\times {\bf R}^4$
coordinates, $f$ is a function which can be found in \BlauNE\ and 
$\epsilon_0$ is a constant $SO(8)$ spinor. The amount of
unbroken supersymmetry is the number of Killing spinors \spinors\ left
invariant under the $Z_2$ action. Therefore, the unbroken
supersymmetries satisfy
\eqn\unbroken{\eqalign{
\epsilon(x,y,x^+)=&g\cdot
\epsilon(x,y,x^+)\cr=&\gamma^{5678}\epsilon(x,-y,x^+)\cr
=&f(x,y,x^+)
\gamma^{5678}\epsilon_0,}}
namely, 
\eqn\constr{
\gamma^{5678}\epsilon_0=\epsilon_0.}
Therefore the orbifold preserves 1/2 of the supersymmetry which is
generated by Killing spinors satisfying this condition.

\newsec{Discussion}

In this paper we have given an explicit example 
of an ${\cal N}=1$ superconformal field theory 
which, in the large $N$ limit, has a subsector 
of the Hilbert space
with enhanced ${\cal N}=4$ superconformal symmetry. 
We have arrived at this perhaps unexpected conclusion 
by taking the corresponding limit in the string theory side 
and by showing that it becomes identical to the theory
with higher supersymmetry. The subsector of the gauge
theory that should exhibit this symmetry enhancement 
is dictated by
the Penrose limit which restricts the space of 
states of the gauge
theory to those whose conformal dimension 
and R-charge diverge in the
large $N$ limit but which nevertheless have 
finite $(\Delta-{3\over 2}R)$.

The light-cone Hamiltonian for the background that one 
obtains in the limit can be
found from the Hamiltonian of \MetsaevBJ\ 
for the maximally supersymmetric case
by twisting it with $U(1)$ charges. In this way 
we get a prediction for the
spectrum of $(\Delta-{3\over 2}R)$ 
of the ${\cal N}=1$ superconformal field theory. 
We proposed how stringy excitations are
related to gauge theory operators 
and made predictions about the gauge theory spectrum.
Perhaps the most striking one is that the various ${\cal N}=1$ multiplets 
should turn into multiplets of ${\cal N}=4$ supersymmetry.
In particular, the chiral multiplets and the semi-conserved 
multiplets of ${\cal N}=1$ supersymmetry should combine into 
${\cal N}=4$ chiral multiplets.
It would be interesting to explore this prediction further. The
duality also predicts values of $(\Delta-{3\over 2}R)$ for certain
operators in this strongly coupled gauge theory.

We have shown that the enhancement of symmetry in the
Penrose limit is a fairly generic phenomenon in theories 
which have gravity duals. One can intuitively understand this
as due to the fact that the limit flattens out
parts of spacetime by focusing on a region near a null
geodesic. For example, we found that the Penrose limit
of a collection of flat NS-NS five-branes is the Nappi-Witten geometry.
The limit of the Maldacena-Nu\~nez geometry is its variation
and also has 16 supercharges. This is an interesting case since we can
quantize the worldsheet theory without taking the light-cone
gauge, and we can compute correlation functions and other
observables using the standard techniques of conformal
field theory \ref\unpublished{J.~Gomis, H.~Ooguri, and E.~Witten,
{\it unpublished.}}.   
On the other hand, we also found cases in which 
the Penrose limit does not lead to symmetry enhancement. 
It would be interesting to explore further the
consequences of this enhancement of symmetries 
for QCD-like theories
and see which lessons this might teach us for 
the familiar questions about
strongly coupled dynamics of these gauge theories.

\bigskip
\bigskip

\centerline{{\bf Acknowledgments }}

\bigskip
We would like to thank Andreas Brandhuber, Sergio Ferrara,
Juan Maldacena, Sunil Mukhi, Horatiu Nastase, and
John Schwarz for useful discussion. 
This research was supported in part by
DOE grant DE-FG03-92-ER40701 and by Caltech Discovery Fund.

\vfill

\appendix{A}{String spectrum in the original coordinates}

It may be instructive to show how we can quantize string theory 
using the metric \ppwave\  before we make
the coordinate transformation to its manifestly symmetric
form \ppmetric. We can read off the bosonic part of the 
light-cone action from the metric \ppwave\ as
\eqn\hamil{\eqalign{
S=& {1\over 2\pi \alpha'}
\int d\tau \int_0^{2\pi \alpha' p^+}d\sigma
\left[ {1\over 2}\sum_{i=1}^4 \left( \dot r_i^2 - r_i'^2 \right)\right.\cr
&~~~~~~ +\left. \sum_{a=1,2} \left( {1\over 2}
 \left( \dot x_a^2  + \dot y_a^2 - x_a'^2 -y_a'^2 \right)
- x_a \dot y_a + y_a \dot x_a \right) \right],}}
where $\dot {} = \partial_\tau$ and $' = \partial_\sigma$. 

The spectrum of the $r_i$ part of the light-cone Hamiltonian
is 
\eqn\rpart{ H_{r-part} = \sum_{n=-\infty}^\infty N_n^{(r)} 
\sqrt{1 + \left({n \over \alpha' p^+ }\right)^2},}
as in the case of the ${\cal N}=4$ theory in \BerensteinJQ ,
where $n$ is the label of the Fourier mode around the $\sigma$
direction and $N_n^{(r)}$ is the number of excitations of that mode. 

Let us examine the $(x_a, y_a)$ part of the Hamiltonian.
In the following, we will ignore the index $a$
with the assumption that we are referring to either $a=1$
or $2$. Accordingly the manifest global symmetry is $U(1)
\subset SU(2)$. In the Fourier expansion,
\eqn\fourierx{ x= {1 \over \sqrt{p^+}}
\sum_n x_n e^{i{n \over \alpha' p^+} \sigma}, ~~
y= {1 \over \sqrt{p^+}}
\sum_n y_n e^{i{n \over \alpha' p^+} \sigma},}
the Hamiltonian becomes
\eqn\xypart{ H_{xy-part} =
\sum_{n=0}^\infty H_n,}
where
\eqn\hn{ \eqalign{ H_{n\neq 0} = & 
\left( p_{x_n} - y_{-n} \right)
\left( p_{x_{-n}} - y_n \right)
+ \left( p_{y_n} + x_{-n} \right)
\left( p_{y_{-n}} + x_n \right)  \cr
&~~~ + \left( {n \over \alpha'
p^+}\right)^2 \left( x_n x_{-n} + y_n y_{-n} \right),\cr }}
and 
\eqn\hzero{ H_0 = {1 \over 2} \left(p_{x_0} - y_0\right)^2
 + {1 \over 2} \left( p_{y_0} + x_0 \right)^2,}
where
\eqn\whatp{ p_{x_n} = -i {\partial \over \partial x_n},~~
p_{y_n} = -i {\partial \over \partial y_n}.}
To compare with the gauge theory spectrum, it is
useful to use the complex coordinates, $z_n = x_n + iy_n$
and $\bar z_n = x_n - i y_n$, so that the $U(1)$ part of
the global $SU(2)$ symmetry is manifest. 
The Hamiltonians for the Fourier modes then become
\eqn\hnfourie{\eqalign{ H_{n\neq 0} =& 
\left( p_{\bar{z}_{-n}} + {i \over 2} z_n\right)
\left( p_{z_n} -{i \over 2} \bar z_{-n} \right)
+ 
\left( p_{\bar{z}_{n}} - {i \over 2} z_{-n}\right)
\left( p_{z_{-n}} + {i \over 2} \bar z_{n} \right) \cr
&~~~~~~+{1 \over 2} \left( {n \over \alpha' p^+}\right)^2 
\left( z_n \bar z_{-n} + z_{-n} \bar z_{n} \right),}}
and
\eqn\hzerofourier{
H_{0} = 
\left( p_{\bar{z}_{0}} + {i \over 2} z_0\right)
\left( p_{z_0} -{i \over 2} \bar z_{0} \right).}
Let us diagonalize them. 

\medskip

\noindent
(a) Zero mode

The Hamiltonian $H_0$ for the zero mode is nothing but the
one for a charged particle in two dimensions in a constant
magnetic field. It has the Landau spectrum with
infinite degeneracy at each level.  To compare with the gauge theory
spectrum, it is useful 
to introduce the following set of oscillators,
\eqn\zeromodeosc{
\eqalign{ &a_0 = {1 \over 2} \bar z_0 + {\partial 
\over \partial z_0},~~a_0^\dagger = 
{1 \over 2} z_0 - {\partial \over \partial \bar z_0}, \cr
&b_0 = {1 \over 2} z_0 + {\partial \over \partial \bar z_0},~~
b_0^\dagger = {1 \over 2} \bar z_0
- {\partial \over \partial z_0}.}}
They obey the commutation relations
\eqn\zerocommutation{ [a_0, a_0^\dagger ] = 1,~~
[b_0, b_0^\dagger ] = 1,~~({\rm others}) = 0.}
Under
the $U(1)$ subgroup of the $SU(2)$ global symmetry,
the operators $a_0, b_0^\dagger$ carry charge $-1$
and $a_0^\dagger, b_0$ carry charge $+1$ 

In terms of the oscillators, 
the Hamiltonian \hzerofourier\ is expressed
as
\eqn\hzeroosc{ H_0 = 2 a_0^\dagger a_0 + 1.}
In particular, it commutes with the oscillators $b_0, b_0^\dagger$.
The lowest energy states are annihilated by 
$a_0$ and thus have the form $(b_0^\dagger)^k \exp(-{1\over 2}z_0 
\bar z_0)$ where $k$ is any non-negative integer.
The complete energy eigenstates are given by
\eqn\zeromodeeigen{
 \psi_{m,k}^{(0)} = 
(a_0^\dagger)^m (b_0^\dagger)^k \exp\left(
-{1\over 2} z_0 \bar z_0 \right),}
with $m, k \geq 0$. The energy of the state is $2m$
and the $U(1)$ global charge is $(k-m)$.

\medskip
\noindent
(b) Non-zero modes

As in the case of the zero mode, we introduce the
following set of oscillators,
\eqn\nmodeosc{
\eqalign{ &a_n = {1 \over 2} \bar z_n + {\partial 
\over \partial z_{-n}},~~a_n^\dagger = 
{1 \over 2} z_{-n} - {\partial \over \partial \bar z_n}, \cr
&b_n = {1 \over 2} z_n + {\partial \over \partial \bar z_{-n}},~~
b_n^\dagger = {1 \over 2} \bar z_{-n} - {\partial \over
\partial  z_n},}}
obeying the commutation relations,
\eqn\nonzerocomm{ [a_n, a_n^\dagger]=1, ~~~[b_n, b_n^\dagger]=1,~~~
({\rm others}) = 0.}
In the limit $n^2/\alpha' p^+ \rightarrow 0$, 
the Hamiltonian $H_n$ takes the same form as in the
case of the zero mode,
\eqn\hnlimit{ H_n = 2a_n^\dagger a_n + 2a_{-n}^\dagger a_{-n} + 2.}
In particular,
each energy level has infinite degeneracy generated
by the oscillators $b_{\pm n}^\dagger$. 
                          
For finite $n^2/\alpha'p^+$, the Hamiltonian $H_n$ contains
terms mixing the two oscillators,
\eqn\hnosc{\eqalign{ H_n = & 2a_n^\dagger a_n
+ {1\over 2} \omega^2 (a_n^\dagger + b_{-n})(a_n + b_{-n}^\dagger)\cr
&~~~~~+(n \rightarrow -n) + 2 ,}}
where we introduced $\omega = n/\alpha'p^+$ to simply the
following equations.  
The Hamiltonian can be diagonalized by introducing
a new set of oscillators $\alpha_n$ and $\beta_n$ defined by
\eqn\reshuffle{
\eqalign{ a_n & = \cosh \theta \alpha_n + \sinh\theta
\beta_{-n}^\dagger, ~~
a_n^\dagger =  \cosh \theta \alpha_n^\dagger + \sinh\theta
\beta_{-n}, \cr
b_n & = \cosh \theta \beta_n + \sinh \theta \alpha_{-n}^\dagger,
~~b_n^\dagger = \cosh \theta \beta_n^\dagger
+\sinh \theta \alpha_{-n}.}}
They obey the commutation relations
\eqn\commalpha{ [\alpha_n, \alpha_n^\dagger] = 1,~~
[\beta_n , \beta_n^\dagger ] = 1, ~~({\rm others})=0.}
The vacuum state for $\alpha, \beta$ is related to the
one for $a, b$ by the Bogolubov transformation. 
Substituting these into \hnosc\ and requiring that
the cross terms of $\alpha$ and $\beta$ to vanish,
we find
\eqn\whattheta{ e^{-4\theta} = 1+\omega^2.}
In terms of the new set of
oscillators, the Hamiltonian $H_n$ is then 
expressed as
\eqn\newhamilton{
 \eqalign{ H_n = & \left( \sqrt{1+\omega^2}+1\right) \alpha_n^\dagger
\alpha_n  + \left(\sqrt{1+\omega^2}-1\right) \beta_n^\dagger\beta_n
\cr
&~~~+(n \rightarrow -n) + 2 \sqrt{1+\omega^2}.}} 
Here we have chosen a solution so that, in the limit
$\omega \rightarrow 0$, the oscillators become $\alpha_n \rightarrow a_n$
and $\beta_n \rightarrow b_n$. 
Energy eigenstates of $H_n$ are then given by
$(\alpha_n^\dagger)^m (\alpha_{-n}^\dagger)^{m'}
(\beta_n^\dagger)^k (\beta_{-n}^\dagger)^{k'} |0\rangle ,$
where  $|0\rangle$ is the vacuum state for the $\alpha, \beta$
oscillators. The state carries the energy, 
$$ E_{m,m',k,k'}= (m+m')  \left( \sqrt{1+\omega^2}+1\right)
+(k+k') \left( \sqrt{1+\omega^2}-1\right),$$
and the global U(1) charge $(k+k'-m-m')$. 

Combining this with the result for the zero mode,
 the complete spectrum for the $(x, y)$ part of the light-cone
Hamiltonian is given by
\eqn\xyspectrum{
  H_{xy-part} = \sum_{n=-\infty}^\infty
\left[  N_n^{(\alpha)} \left( \sqrt{
1 +\left({n\over \alpha' p^+}\right)^2}+1 \right) 
+  N_n^{(\beta)}\left( \sqrt{
1 +\left({n\over \alpha' p^+}\right)^2}-1 \right) \right].}
This result agrees with the one we obtained in section 2, 
with the identification:
\eqn\identification{
 \alpha_n = \bar w_n,~~ \beta_n = w_n}
in \contri .

\listrefs

\end